\documentclass[12pt]{article}
\usepackage{graphicx}
\def \b{{\cal B}}
\def \bea{\begin{eqnarray}}
\def \beq{\begin{equation}}

\def \eea{\end{eqnarray}}
\def \eeq{\end{equation}}

\begin{document}
\Large
\centerline {\bf Prospects for detection of $\Upsilon(1D) \to \Upsilon(1S)
\pi \pi$}
\centerline{\bf via $\Upsilon(3S) \to \Upsilon(1D) + X$ 
\footnote{Enrico Fermi Institute preprint EFI 03-06, hep-ph/0302122.
To be submitted to Physical Review D, Brief Reports.}}
\normalsize
\bigskip
 
\centerline{Jonathan L. Rosner~\footnote{rosner@hep.uchicago.edu}}
\centerline {\it Enrico Fermi Institute and Department of Physics}
\centerline{\it University of Chicago, 5640 S. Ellis Avenue, Chicago, IL 60637}
\bigskip
 
\begin{quote}

At least one state in the first family of D-wave $b \bar b$ quarkonium levels
has been discovered near the predicted mass of 10.16 GeV/$c^2$.  This state
is probably the one with $J=2$.  This state and the ones with $J=1$ and $J=3$
may contribute a detectable amount to the decay $\Upsilon(1D) \to \Upsilon(1S)
\pi \pi$, depending on the partial widths for these decays for which
predictions vary considerably.  The prospects for detection of the chain
$\Upsilon(3S) \to \Upsilon(1D) + X \to \Upsilon \pi \pi + X$ are discussed. 

\end{quote}
\bigskip

\noindent
PACS Categories:  14.40.Gx, 13.20.Gd, 13.40.Hq, 12.39.Ki
\bigskip

A sample of $4.73 \times 10^6$ $\Upsilon(3S)$ decays obtained by the CLEO III
experiment has permitted the first observation of an $\Upsilon(1D)$ level
\cite{TS02}.  This state, which probably is the $J=2$ member of the
spin-triplet $1^3D_{1,2,3}$, has a mass of $10162.2 \pm 1.6$ MeV$/c^2$ and
is observed with a combined set of branching ratios for production and decay
very close to theoretical predictions \cite{KR88,GR02}.  The data exceed the
CLEO Collaboration's previously reported sample \cite{Brock91}, $237~000 \pm
6800$ $\Upsilon(3S)$ decays, by about a factor of 20.  They are several times
that of the CUSB Collaboration \cite{Heintz92}, $1.18 \times 10^6$
$\Upsilon(3S)$ decays with $\mu^+ \mu^-$ detection and $0.64 \times 10^6$
decays with $e^+ e^-$ detection.

The chain through which the $\Upsilon(1D)$ state was discovered is illustrated
in Fig.\ 1(a).  A cascade involving four photons $\gamma_1$--$\gamma_4$
followed by leptonic decay of the $\Upsilon(1S)$ leads to a very clean final
state in which the major limitations are the total size of the sample and
backgrounds from cascades via the photons labeled ${\gamma_2}'$ and
${\gamma_3}'$ involving the intermediate $\Upsilon(2S)$ state.
The combined branching ratio including the decay $\Upsilon(1S) \to (e^+ e^-
~{\rm or}~\mu^+ \mu^-)$, $(3.3 \pm 0.6 \pm 0.5) \times 10^{-5}$, is consistent
with the prediction \cite{GR02} of $3.76 \times 10^{-5}$.

\begin{figure}
\begin{center}
\includegraphics[height=3.2in]{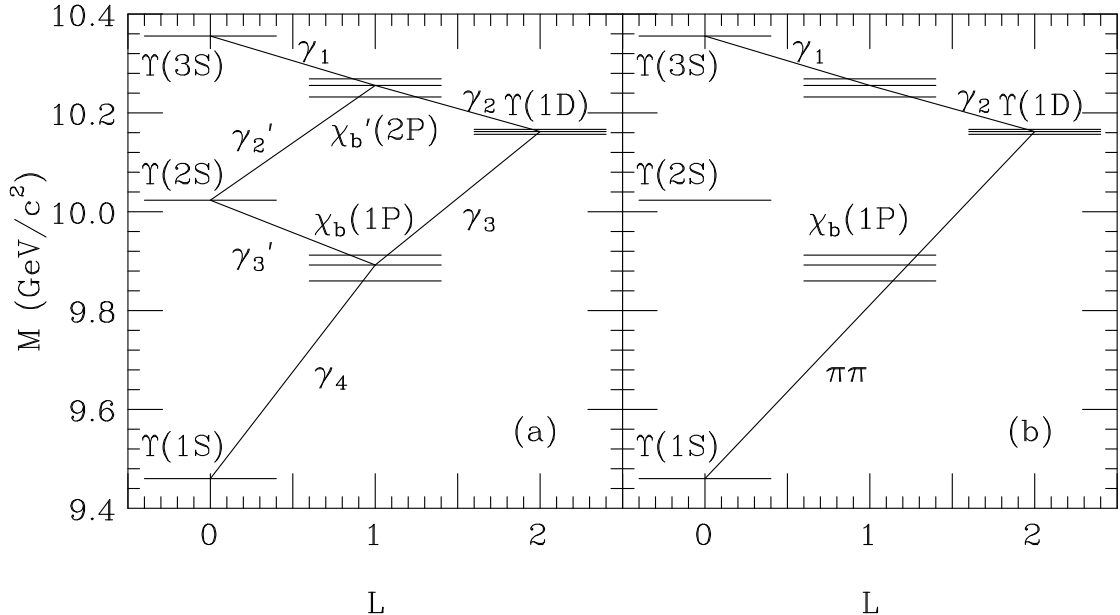}
\caption{Cascades from the $\Upsilon(3S)$ leading to production and detection
of the $\Upsilon(1D)$ levels.  In each case the $\Upsilon(1S)$ is detected
via its decay to $e^+ e^-$ or $\mu^+ \mu^-$.  (a) Four-photon cascades;
(b) sequence $\Upsilon(3S) \to \gamma_1 \gamma_2 \Upsilon(1D)$ followed
by $\Upsilon(1D) \to \Upsilon(1S) \pi \pi$.
\label{fig:trans}}
\end{center}
\end{figure}

Theoretical predictions \cite{GR02} indicate that for the CLEO III
$\Upsilon(3S)$ sample mentioned above, one should have produced a total of
(5.7, 14.2, 7.7) thousand $\Upsilon(1D)$ states with $J = (1,~2,~3)$,
respectively, via the transitions involving $\gamma_1$ and $\gamma_2$ in Fig.\
1(a).  Yan \cite{Yan80} and Tuan \cite{Tuan96} have pointed out that another
way to detect these states may be through their decays to $\Upsilon(1S) \pi
\pi$, as illustrated in Fig.\ 1(b).  The present note examines the likelihood
of observing these transitions in the CLEO III data set.  It is not necesssary
to detect the photons $\gamma_1$ and $\gamma_2$, though vetoing charged
particles aside from the pions in $\Upsilon(1D) \to \Upsilon(1S) \pi \pi$
might be helpful.

The predictions for $\Upsilon(1D) \to \Upsilon(1S) \pi \pi$ rates differ
substantially from one another.  All are independent of the spin of the $1D$
state \cite{Yan80,Mox88}, aside from possible small effects on the $J=1$ state
due to $^3S_1$--$^3D_1$ mixing \cite{Kuang}. Billoire {\it et al.}
\cite{BLMN79} found $\Gamma
(\Upsilon(1D) \to \Upsilon(1S) \pi \pi) = 0.07 \alpha_s^2$ keV; Kuang and Yan
\cite{KY} obtained 24 KeV; Moxhay \cite{Mox88} found 0.07 keV (a value employed
in Refs.\ \cite{KR88} and \cite{GR02}), and Ko \cite{Ko93} argued
on the basis of non-perturbative effects that the Billoire {\it et al.} and
Moxhay results should be enhanced, leading to a prediction of $0.56 \pm 0.07$
keV.  I shall compare the sensitivity of the present CLEO III sample to
the last three predictions.  Discussions of the reasons for the differences
may be found in Refs.\ \cite{KR88,Ko93,Vol86,KY90}.

In brief, the present data sample should be able to rule out the
prediction of Kuang and Yan or observe the predicted transition with great
statistical significance.  The sensitivity at the level of Ko's prediction will
depend on CLEO III's superior performance as a detector in comparison with the
previous CLEO limit \cite{Brock91}.  In that work a combined upper limit 
\beq \label{eqn:oldCLEO}
\sum_{J=1}^3 \b[\Upsilon(3S) \to \Upsilon(1^3D_J) X]
\b[\Upsilon(1^3D_J \to \Upsilon(1S) \pi \pi] < 0.6\%~(90\%~{\rm c.l.})
\eeq
was placed.

I first justify the estimate of the total number of $\Upsilon(1D)$ states
produced from the $\Upsilon(3S)$ through two-photon cascades via the
${\chi_b}'(2P)$ states.  The branching ratios for the transitions $\Upsilon(3S)
\to {\chi_b}'(2P) \gamma_1$ are quoted in Ref.\ \cite{PDG} (central
values are used here), while for the decays ${\chi_b}'(2P) \to \Upsilon(1D)
\gamma_2$ the branching ratios predicted in Table VIII of Ref.\ \cite{KR88}
are employed.  The combined results are summarized in Table I.  It is assumed
that these electromagnetic transitions are the only source of the
$\Upsilon(1D)$ levels.  The energies of the photons $\gamma_1$ are shown in the
Table, while the energies of $\gamma_2$ range between about 80 and 120 MeV for
the masses favored in Ref.\ \cite{KR88}, shown in parentheses in MeV/$c^2$
under the corresponding spins of the $1^3D_J$ levels.

\begin{table}
\caption{Branching ratios for electromagnetic transitions contributing to
$\Upsilon(3S) \to \Upsilon(1^3D_J) X$.  Branching ratios for emission of
$\gamma_1$ are observed values, taken from Ref.\ \cite{PDG}.  Table entries in
Roman type denote $\b({\chi_{bJ}}' \to \Upsilon(1^3D_J) \gamma_2)$ predictions
in percent \cite{KR88}; numbers in italics denote thousands of
$\Upsilon(1^3D_J)$ states produced for $4.73 \times 10^6$ $\Upsilon(3S)$
decays.  Blank entries denote transitions forbidden by electric dipole
selection rules.
\label{tab:brstoD}}
\begin{center}
\begin{tabular}{c c c c c c} \hline \hline
 & $J({\chi_b}')$     &        0      &         1      & 2 & Total       \\
 & $\b(\gamma_1)(\%)$ & $5.4 \pm 0.6$ & $11.3 \pm 0.6$ & $11.4 \pm 0.8$ &
events \\
 & $E_{\gamma_1}$ (MeV) &    123      &      100     & 87 & ($\times 10^3)$ \\
\hline
$^3D_1$ &               &   0.153     &     0.97     & 0.023 &     \\
(10156) &            & {\it 0.391}   & {\it 5.185} & {\it 0.124} & {\it 5.70}\\
$^3D_2$ &               &             &     2.36     & 0.29  &      \\
(10162) &            &          & {\it 12.61} & {\it 1.564} & {\it 14.18} \\
$^3D_3$ &               &             &              & 1.42 &     \\
(10166) &           &            &             & {\it 7.657} & {\it 7.66} \\
\hline \hline
\end{tabular}
\end{center}
\end{table}

To calculate the branching ratios of each of the $1^3D_J$ states into
$\Upsilon(1S) \pi \pi$ for the various models of rates, one may use the
predictions \cite{KR88} of decay rates of each state into other modes,
$\Gamma_{\rm other}(1^3D_{1,2,3}) = (35.5,~28.0,~25.4)$ keV.
The results are summarized in Table \ref{tab:Dbrs}, taking
Ko's central value of $\Gamma(1^3D_J \to \Upsilon(1S) \pi \pi) = 0.56$ keV.

\begin{table}
\caption{Predicted branching ratios for $\Upsilon(1^3D_J) \to \Upsilon(1S) \pi
\pi$ (Roman entries, in percent) and numbers of $\Upsilon(1S) \pi \pi$ final
states produced for $4.73 \times 10^6$ $\Upsilon(3S)$ decays (italics).
\label{tab:Dbrs}}
\begin{center}
\begin{tabular}{c c c c c} \hline \hline
State & $N(^3D_J)$    & Kuang-Yan &    Moxhay    &       Ko    \\
    & $(\times 10^3)$ & \cite{KY} & \cite{Mox88} & \cite{Ko93} \\ \hline        
$1^3D_1$ &            &   40.4    &    0.20      &     1.55    \\
         & 5.70      & {\it 2303} &  {\it 11}    & {\it 88}   \\
$1^3D_2$ &            &   46.2    &    0.25      &     1.96    \\
         & 14.18     & {\it 6551} &  {\it 35}    & {\it 278}   \\
$1^3D_3$ &            &   48.6    &    0.27      &     2.16    \\
         & 7.66      & {\it 3721} &  {\it 21}    & {\it 165}   \\
Total    & 27.54    & {\it 12575} &  {\it 67}    & {\it 532}   \\
\multicolumn{2}{c}{$\b$ [Eq.\ (\ref{eqn:oldCLEO}) (\%)]} & 0.27 &
0.0014 & 0.011 \\ \hline \hline
\end{tabular}
\end{center}
\end{table}

The predicted products of branching ratios corresponding to the left-hand side
of Eq.\ (\ref{eqn:oldCLEO}) are all below the old CLEO upper bound
\cite{Brock91}.  However, with 20 times the old data, CLEO III should be able
to check the Kuang--Yan prediction.  (Our value of 0.27\% is to be compared
with the estimate made by Tuan \cite{Tuan96} of 0.43\%).  The prediction of
Moxhay appears to be too low to be testable.  Ko \cite{Ko93} has argued that
nonperturbative effects raise Moxhay's prediction by a factor of $8 \pm 1$,
leading to the prediction
\beq \label{eqn:Ko}
\sum_{J=1}^3 \b[\Upsilon(3S) \to \Upsilon(1^3D_J) X]
\b[\Upsilon(1^3D_J \to \Upsilon(1S) \pi \pi] = 0.011\%~~.
\eeq
(Ko obtained a slightly smaller value of 0.009\%;
Eq.\ (\ref{eqn:Ko}) takes account of updated estimates \cite{GR02} of
$\Upsilon(1^3D_J)$ production.) Although this is a factor of about 50 below the
upper bound in Eq.\ (\ref{eqn:oldCLEO}), the total number of predicted events
looks encouraging.  Multiplying 532 $\Upsilon(1S) \pi \pi$ decays by 2/3
(for charged pions) and by $\b[\Upsilon(1S) \to (e^+ e^-~{\rm or}~\mu^+ \mu^-)]
= 4.86 \pm 0.13\%$, we expect about 17 events with two charged pions and two
charged leptons ($e$ or $\mu$ pairs).  Of these, about half should be due to
the $1^3D_2$ state which is the best candidate for that reported in Ref.\
\cite{TS02}, with about 2/3 of the remaining events due to the $1^3D_3$ state.
Some additional sensitivity might be gained using neutral pions.

\bigskip
I am grateful to Stephen Godfrey, Yu-Ping Kuang, San Fu Tuan, and
Eckhard von Toerne for
discussions.  This work was supported in part by the United States Department
of Energy through Grant No.\ DE FG02 90ER40560.

\def \ajp#1#2#3{Am.\ J. Phys.\ {\bf#1}, #2 (#3)}
\def \apny#1#2#3{Ann.\ Phys.\ (N.Y.) {\bf#1}, #2 (#3)}
\def \app#1#2#3{Acta Phys.\ Polonica {\bf#1}, #2 (#3)}
\def \arnps#1#2#3{Ann.\ Rev.\ Nucl.\ Part.\ Sci.\ {\bf#1}, #2 (#3)}
\def \art{and references therein}
\def \cmts#1#2#3{Comments on Nucl.\ Part.\ Phys.\ {\bf#1}, #2 (#3)}
\def \cn{Collaboration}
\def \cp89{{\it CP Violation,} edited by C. Jarlskog (World Scientific,
Singapore, 1989)}
\def \efi{Enrico Fermi Institute Report No.\ }
\def \epjc#1#2#3{Eur.\ Phys.\ J. C {\bf#1}, #2 (#3)}
\def \f79{{\it Proceedings of the 1979 International Symposium on Lepton and
Photon Interactions at High Energies,} Fermilab, August 23-29, 1979, ed. by
T. B. W. Kirk and H. D. I. Abarbanel (Fermi National Accelerator Laboratory,
Batavia, IL, 1979}
\def \hb87{{\it Proceeding of the 1987 International Symposium on Lepton and
Photon Interactions at High Energies,} Hamburg, 1987, ed. by W. Bartel
and R. R\"uckl (Nucl.\ Phys.\ B, Proc.\ Suppl., vol.\ 3) (North-Holland,
Amsterdam, 1988)}
\def \ib{{\it ibid.}~}
\def \ibj#1#2#3{~{\bf#1}, #2 (#3)}
\def \ichep72{{\it Proceedings of the XVI International Conference on High
Energy Physics}, Chicago and Batavia, Illinois, Sept. 6 -- 13, 1972,
edited by J. D. Jackson, A. Roberts, and R. Donaldson (Fermilab, Batavia,
IL, 1972)}
\def \ijmpa#1#2#3{Int.\ J.\ Mod.\ Phys.\ A {\bf#1}, #2 (#3)}
\def \ite{{\it et al.}}
\def \jhep#1#2#3{JHEP {\bf#1}, #2 (#3)}
\def \jpb#1#2#3{J.\ Phys.\ B {\bf#1}, #2 (#3)}
\def \lg{{\it Proceedings of the XIXth International Symposium on
Lepton and Photon Interactions,} Stanford, California, August 9--14 1999,
edited by J. Jaros and M. Peskin (World Scientific, Singapore, 2000)}
\def \lkl87{{\it Selected Topics in Electroweak Interactions} (Proceedings of
the Second Lake Louise Institute on New Frontiers in Particle Physics, 15 --
21 February, 1987), edited by J. M. Cameron \ite~(World Scientific, Singapore,
1987)}
\def \kdvs#1#2#3{{Kong.\ Danske Vid.\ Selsk., Matt-fys.\ Medd.} {\bf #1},
No.\ #2 (#3)}
\def \ky85{{\it Proceedings of the International Symposium on Lepton and
Photon Interactions at High Energy,} Kyoto, Aug.~19-24, 1985, edited by M.
Konuma and K. Takahashi (Kyoto Univ., Kyoto, 1985)}
\def \mpla#1#2#3{Mod.\ Phys.\ Lett.\ A {\bf#1}, #2 (#3)}
\def \nat#1#2#3{Nature {\bf#1}, #2 (#3)}
\def \nc#1#2#3{Nuovo Cim.\ {\bf#1}, #2 (#3)}
\def \nima#1#2#3{Nucl.\ Instr.\ Meth. A {\bf#1}, #2 (#3)}
\def \np#1#2#3{Nucl.\ Phys.\ {\bf#1}, #2 (#3)}
\def \npbps#1#2#3{Nucl.\ Phys.\ B Proc.\ Suppl.\ {\bf#1}, #2 (#3)}
\def \os{XXX International Conference on High Energy Physics, Osaka, Japan,
July 27 -- August 2, 2000}
\def \PDG{Particle Data Group, K. Hagiwara \ite, \prd{66}{010001}{2002}}
\def \pisma#1#2#3#4{Pis'ma Zh.\ Eksp.\ Teor.\ Fiz.\ {\bf#1}, #2 (#3) [JETP
Lett.\ {\bf#1}, #4 (#3)]}
\def \pl#1#2#3{Phys.\ Lett.\ {\bf#1}, #2 (#3)}
\def \pla#1#2#3{Phys.\ Lett.\ A {\bf#1}, #2 (#3)}
\def \plb#1#2#3{Phys.\ Lett.\ B {\bf#1}, #2 (#3)}
\def \pr#1#2#3{Phys.\ Rev.\ {\bf#1}, #2 (#3)}
\def \prc#1#2#3{Phys.\ Rev.\ C {\bf#1}, #2 (#3)}
\def \prd#1#2#3{Phys.\ Rev.\ D {\bf#1}, #2 (#3)}
\def \prl#1#2#3{Phys.\ Rev.\ Lett.\ {\bf#1}, #2 (#3)}
\def \prp#1#2#3{Phys.\ Rep.\ {\bf#1}, #2 (#3)}
\def \ptp#1#2#3{Prog.\ Theor.\ Phys.\ {\bf#1}, #2 (#3)}
\def \rmp#1#2#3{Rev.\ Mod.\ Phys.\ {\bf#1}, #2 (#3)}
\def \rp#1{~~~~~\ldots\ldots{\rm rp~}{#1}~~~~~}
\def \rpp#1#2#3{Rep.\ Prog.\ Phys.\ {\bf#1}, #2 (#3)}
\def \sing{{\it Proceedings of the 25th International Conference on High Energy
Physics, Singapore, Aug. 2--8, 1990}, edited by. K. K. Phua and Y. Yamaguchi
(Southeast Asia Physics Association, 1991)}
\def \slc87{{\it Proceedings of the Salt Lake City Meeting} (Division of
Particles and Fields, American Physical Society, Salt Lake City, Utah, 1987),
ed. by C. DeTar and J. S. Ball (World Scientific, Singapore, 1987)}
\def \slac89{{\it Proceedings of the XIVth International Symposium on
Lepton and Photon Interactions,} Stanford, California, 1989, edited by M.
Riordan (World Scientific, Singapore, 1990)}
\def \smass82{{\it Proceedings of the 1982 DPF Summer Study on Elementary
Particle Physics and Future Facilities}, Snowmass, Colorado, edited by R.
Donaldson, R. Gustafson, and F. Paige (World Scientific, Singapore, 1982)}
\def \smass90{{\it Research Directions for the Decade} (Proceedings of the
1990 Summer Study on High Energy Physics, June 25--July 13, Snowmass, Colorado),
edited by E. L. Berger (World Scientific, Singapore, 1992)}
\def \tasi{{\it Testing the Standard Model} (Proceedings of the 1990
Theoretical Advanced Study Institute in Elementary Particle Physics, Boulder,
Colorado, 3--27 June, 1990), edited by M. Cveti\v{c} and P. Langacker
(World Scientific, Singapore, 1991)}
\def \yaf#1#2#3#4{Yad.\ Fiz.\ {\bf#1}, #2 (#3) [Sov.\ J.\ Nucl.\ Phys.\
{\bf #1}, #4 (#3)]}
\def \zhetf#1#2#3#4#5#6{Zh.\ Eksp.\ Teor.\ Fiz.\ {\bf #1}, #2 (#3) [Sov.\
Phys.\ - JETP {\bf #4}, #5 (#6)]}
\def \zpc#1#2#3{Zeit.\ Phys.\ C {\bf#1}, #2 (#3)}
\def \zpd#1#2#3{Zeit.\ Phys.\ D {\bf#1}, #2 (#3)}

\end{document}